# Interlocking mechanism between molecular gears attached to surfaces

Rundong Zhao[1], Yan-Ling Zhao[2], Fei Qi[1], Klaus Hermann[3], Rui-Qin Zhang[2,4] and Michel A. Van Hove[1*]

[1]Institute of Computational and Theoretical Studies & Department of Physics, Hong Kong Baptist University, Hong Kong SAR, China
[2]Department of Physics and Materials Science, City University of Hong Kong, Hong Kong SAR, China
[3]Inorganic Chemistry Department, Fritz-Haber-Institute der Max-Planck-Gesellschaft, Berlin, Germany
[4]Beijing Computational Science Research Center, Beijing, China

* Corresponding author; E-mail: vanhove@hkbu.edu.hk

## Abstract

While molecular machines play an increasingly significant role in nanoscience research and applications, there remains a shortage of investigations and understanding of the molecular gear (cogwheel), which is an indispensable and fundamental component to drive a larger correlated molecular machine system. Employing ab initio calculations, we investigate model systems consisting of molecules adsorbed on metal or graphene surfaces, ranging from very simple triple-arm gears such as $PF_3$ and $NH_3$ to larger multi-arm gears based on carbon rings. We explore in detail the transmission of slow rotational motion from one gear to the next by these relatively simple molecules, so as to isolate and reveal the mechanisms of the relevant intermolecular interactions. Several characteristics of molecular gears are discussed, in particular the flexibility of the arms and the slipping and skipping between interlocking arms of adjacent gears, which differ from familiar macroscopic rigid gears. The underlying theoretical concepts suggest strongly that other analogous structures may also exhibit similar behavior which may inspire future exploration in designing large correlated molecular machines.

## Introduction

The past twenty years have seen an extraordinary evolution in the development of molecular motors, ranging from single rotating or reciprocating molecules to networks of many identical such molecules.(*1-9*) Some of these molecules are powered by light(*1, 2, 10*), others by chemical reactions(*11-13*), by STM tips(*5, 6*), by electric currents(*14*), or by electric fields(*15, 16*). Individual molecular rotors are already fairly well understood, but there is a need to go further and explore cooperative motions among an interlocking collection of such molecules, analogous to e.g. conventional macroscopic machines and natural muscles that consist of multiple mechanically linked smaller "machines". In particular, the rational design of molecular gears becomes necessary to transmit the rotational motion from its source to a point of application, as in macroscopic machines. The transmission mechanisms of rotation from one gear to the next have not yet been studied on surfaces and is the focus of this paper.

Also important is the ability to combine the mechanical output of many single molecular machines in order to magnify their effect: This requires a network of mechanical linkages to funnel the combined mechanical motion to the desired application. Such mechanical linkages can also greatly help overcome the challenge of thermal randomness: thermal energy will make some molecular machines rotate in the direction opposite to what is desired, which is especially likely at the nanoscale since their rotational and vibrational degrees of freedom are closely and inevitably coupled. Mechanically linking multiple molecular machines will significantly help counteract this randomization by ensuring that "the majority wins" and that the combined unidirectional motion is still a large magnification of the individual motions. Of great interest is furthermore how collective motions propagate across a one-dimensional chain or a multi-dimensional array of rotor molecules, especially in the presence of defects, dissipation, etc., which will be largely inevitable in practical machinery.

Despite experimental challenges, some molecular gears(*7, 17*) and connected rotors(*8, 18, 19*) have already been reported in the literature. Most of these are relatively large single molecules that form networks on metal surfaces by self-assembly. The present work aims to design smaller and simpler molecular gear systems that can help realize the aforementioned functions on surfaces while more clearly exhibiting their underlying mechanisms. Surfaces are often taken as supports of molecular machines because they can keep such machines oriented and fixed in place, while the machines can be individually excited and imaged by surface science techniques such as Scanning Tunneling Microscopy (STM) and Atomic Force Microscopy (AFM). Many surface-adsorbed molecular motors reported in the literature are designed to work on surfaces of metals such as copper and gold. Further, graphene or highly ordered pyrolytic graphite (HOPG) are also suitable substrates for attaching molecular rotors.(*20*)

Designing a molecular-scale gear is different from designing a hard macroscopic gear since it involves interactions between "soft" gears, which may include electrostatic and van der Waals forces. In the present work, taking stable molecules acting as cogwheels with different numbers of arms (from 3 to 6) on metal and graphene surfaces, we show that, by the rational use of intermolecular interactions, such molecules can form interlocked one dimensional chains of gears that can rotate cooperatively. By modeling a series of gear shapes, gear types and supporting surfaces, we examine in particular how transmission of rotation depends on the intermolecular distance and alignment, the nature of the support, the number of arms and the type of molecular interactions. This allows us to reach general conclusions valid for a wider range of situations. All the molecules studied have the potential to perform as molecular gears and may be used as components of larger coordinated machine systems. In a future publication(*21*), we shall address the more distant propagation of rotation from one "driver" molecule through a chain of multiple "slave" molecules.

Our main focus is on "low-speed" non-vibrational interactions between gears, while we include some limited molecular dynamics calculations to explore temporal effects. The simulations do not include "motor" molecules that convert energy input into driving forces; instead, the rotation will be externally/manually applied to "driver" molecules that can transmit this rotation to "slave" molecules whose behavior is then studied; our "driver" molecules are thus generic (and in an experimental realization the "drivers" might even be different from the "slaves"). As such, we are not attempting to model surface systems that might be realized experimentally exactly as modeled, since we are not including the many complications of a motor molecule: the focus is instead on computationally analyzing the nature and behavior of neighboring gear-gear interactions. Thus, this work is a theoretical investigation which aims to reveal the mechanisms of intermolecular interactions in molecular gears, and to provide a general theoretical foundation for a variety of future experimental designs, not only for one particular experimental realization.

**Computational methods**

Calculations in the present paper were performed within first-principles density functional theory (DFT) applying the generalized gradient approximation (GGA) of Perdew, Burke, and Ernzerhof (PBE)(*22*) together with dispersion corrections according to Grimme(*23*) to account for van der Waals interactions. Geometric structures were optimized using the Vienna Ab-initio Simulation Package (VASP)(*24, 25*) based on a periodic supercell concept and the energy decomposition analysis was performed using the Amsterdam Density Functional package (ADF)(*26*) with the Morokuma energy decomposition scheme(*27*).

To simulate the behavior of rotating $PF_3$ and $NH_3$ molecules on copper surfaces, we used slab models that contain three layers of the copper substrate, with the deepest layer fixed (to model the bulk of the metal) and the two layers closest to the surface allowed to fully relax in the optimizations. This substrate model is realistic since the layer relaxations are found to be small with largest perpendicular displacements below 0.04 Å. For multi-arm molecules on graphene, the single layer of graphene (Gr) was fully relaxed, yielding perpendicular displacements of 0.00-0.50 Å (implying buckling of the graphene sheet). On both Cu and Gr, a repeating 2D surface unit cell contains 2 molecules (one "driver" and one "slave"), as shown for example in Fig. 1. The Brillouin zones were sampled by $3 \times 3 \times 1$ Γ-centered k-points for $PF_3$ and $NH_3$ molecules on copper surfaces; for multi-arm gears built on graphene, sampling by $2 \times 2 \times 1$ k-points is sufficient for the convergence of the structures shown in Fig. 5. For all geometry optimizations, the projector-augmented wave (PAW) method(*28, 29*) implemented in VASP was adopted, with a maximum 400eV cutoff for the plane-wave basis set.

In the molecular dynamics simulations for the $PF_3$ molecular gears, the methods implemented in VASP are applied with canonical ensembles (NVT ensembles, with particle number N, system volume V, and temperature T fixed as constant). Further we set the initial and final temperature to near room temperature (300K), and the initial velocities of the ions are chosen randomly, according to a Maxwell-Boltzmann distribution; in other respects, the scheme of the simulation is totally identical to that used in the DFT calculations. The time steps in the simulations are set to 1 fs with a simulation time of 2 ps.

**Results and discussion**

**Modeling gear interactions.** Using DFT calculations, we study the interactions between molecular gears arranged in infinite linear rows, by imposing successive rotations on the "driver" molecule to obtain resulting forces, torques and rotations on neighboring "slave" molecules. As an example, Figs. 1 and S1 show a linear chain of alternating $AB_3$ driver and slave molecules ($AB_3 = PF_3$ or $NH_3$ in the calculations) on a Cu(111) surface, where each driver molecule is in contact with two slaves, and each slave is in contact with two drivers at inter-gear distances $d_c = \sqrt{3}d$ (*d* is the Cu-Cu nearest neighbor distance at the substrate surface). There is little chance of lateral diffusion of the molecules along the surface, due to diffusion potential barriers of about 0.36 eV vs. rotational barriers of about 0.002 eV for single PF$_3$ molecules.

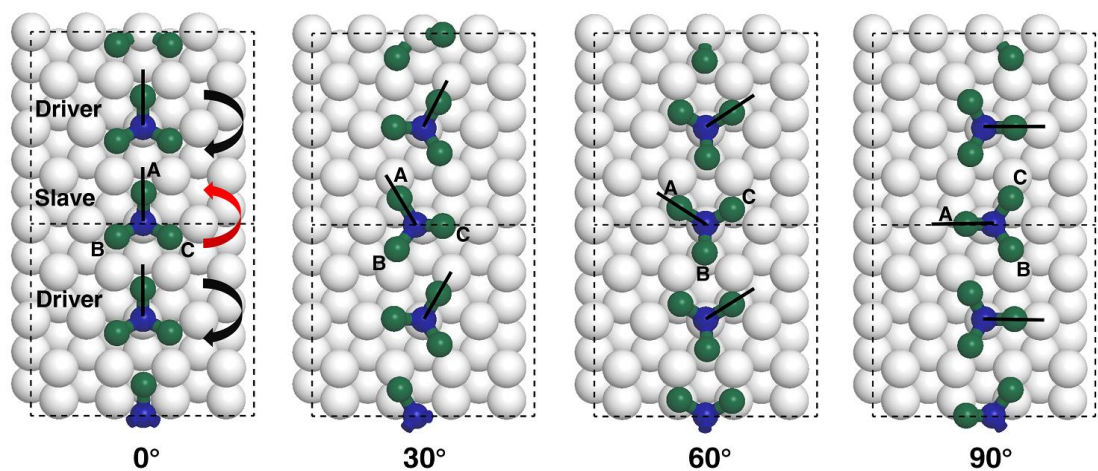

**Fig. 1** Calculated rotation (in top-down views) in a linear gear chain consisting of equidistant alternating driver and slave PF$_3$ molecules (labelled accordingly in the leftmost panel) on Cu(111), arranged in periodic unit cells shown by dashed lines. The PF$_3$ slaves (with arms labelled A, B, C) respond (as indicated by curved red arrows) to the clockwise rotation of their driver PF$_3$ neighbors (indicated by curved black arrows). The four panels correspond to four orientations (marked by black bars) of the driver PF$_3$: 0°, 30°, 60° and 90°, respectively. The inter-gear distance is $d_c = \sqrt{3}d$, where $d$ is the Cu-Cu nearest neighbor distance at the substrate surface.

In our model of molecular rotations, the driver and slave molecules start with realistic, fully optimized adsorption geometries and inter-molecular distances. The drivers are then made to rotate stepwise in a smooth quasi-static motion, to simulate how actual motor molecules might rotate. We allow full relaxation of the slave molecules at each imposed driver rotation angle, such that the slaves can follow any appropriate energy-optimized pathway, including rotation, tilting, bending, slipping, breaking up, etc. (the internal driver structure remains rigid during its imposed rotation, so it can also represent another type of motor molecule with possibly different properties; in future work of longer slave chains(*21*) we allow soft slave-slave interactions and verify that rigid drivers do not constrain our results). We thus ignore dynamic effects such as energy conservation (of kinetic and potential energy), angular momentum conservation and dissipation (energy loss); however, relaxation is allowed to distribute "elastic energy" into the substrate. Each driver thus acts as an externally powered motor that rotates very slowly compared to free-molecule rotations, according to whatever driving mechanism activates it (e.g. slow external forces). It would also be possible to allow fast spontaneous rotations of the drivers (e.g. following a near-instantaneous electronic excitation or collisions inducing rotation); realistically such spontaneous rotations should include energy and angular momentum conservation as well as dissipative effects.

To study the interactions between gears, we start with a high-symmetry gear orientation (0° in Fig. 1): while the drivers are kept aligned and fixed (with the geometry and Cu-P distance obtained from a separate optimization that includes only one PF$_3$ molecule and the Cu substrate), the slaves are fully relaxed together with all Cu atoms except those of the deepest Cu layer (a commonly accepted constraint in modeling molecules on metal surfaces). This minimizes the net forces and torques on all atoms and molecules. Then, in "step 1", we rotate the driver molecules rigidly clockwise by 10° about their molecular rotation axes (which are close to perpendicular to the surface), while keeping the slave molecules fixed: this step allows us to evaluate the forces and torques induced by the drivers on the slaves, as it creates a strained unstable situation. Next, in "step 2", we re-optimize the slave molecules in the presence of the drivers fixed at their new 10° rotated geometry, yielding negligible net forces and torques. This 2-step process is repeated iteratively as we rotate the drivers clockwise by 10° each time until they have reached a rotation angle of 120°, which covers the complete range of rotations, taking the 3-fold symmetry of the fcc (111) substrate into account. The two total energy curves shown in Fig. 2 refer, for each driver rotation angle, to energies obtained after each "step 1" and each "step 2", respectively. Since the single molecules considered here show only small distortions during rotation, the approximation of taking the drivers as rigid bodies is reasonable. However, for a supramolecular system that is very flexible and contains many degrees of rotation and bending, this assumption might significantly restrict the rotational motion pathways and the scenario would often be more complex.(*30*)

Note that the 10° torsion between drivers and slaves breaks the symmetry of the system, so that the plots of forces and torques of Fig. 3 do not exhibit the symmetry of the corresponding equilibrium structures. Section II and Fig. S2 of the Supplementary Information give further details of the iterative 2-step process.

**The gear effect of PF$_3$ and NH$_3$ at the Cu(111) surface.** We first consider a gear molecule PF$_3$ adsorbed at a top site of the Cu(111) surface with its 3-fold axis perpendicular to the substrate. The PF$_3$ molecule is strongly bound to the copper surface due to its lone pair electrons at the phosphorus end. This makes it one of the smallest molecular gears with its three fluorine atoms acting as arms. While it is strongly polarized and can attach to metal surfaces very robustly, the repulsive electrostatic interaction between adjacent PF$_3$ molecules is also strong, providing intense coupling of neighboring gears.

We assemble a linear chain of PF$_3$ molecules as shown in Fig. 1, and then simulate the molecular rotations along the chain following the procedure described above. The distance between adjacent PF$_3$ molecules is set to $d_c = \sqrt{3}d$ ($d \approx 2.556$Å denotes the distance between nearest-neighbor copper atoms at the surface, while the distance $d_c = \sqrt{3}d$ separates identical sites along the vertical axis in Fig. 1): a closer gear-gear distance of $d_c = d$ leads to an unstable configuration in which some of the PF$_3$ molecules will finally leave the surface due to the excessively strong intermolecular repulsion; with a larger distance of e.g. $d_c = 2d$, the interaction between gears becomes too weak to generate effective driving forces: slipping then occurs between drivers and slaves. As a result, Fig. 1 shows the geometry of the PF$_3$ chain for $d_c = \sqrt{3}d$ where the slave molecules are optimized after rotation of the driver molecules by 0° (initial geometry), 30°, 60° and 90°. Clearly, the rotational motion can be successfully transferred from driver (clockwise) to slave (anti-clockwise) with no appreciable angular lag (delay) of rotation and no significant slippage. The slaves are found to slightly tilt off center due to the gear-gear repulsion, which is manifested most notably at the rotation angle of 30° and 90° of the drivers. A complementary molecular dynamics simulation using a canonical ensemble with the temperature fixed at 300K supports this conclusion: after the driver gears are rotated by 10° (and then fixed, as in step 2 of the quasi-static modelling mentioned above), it takes the slave gears less than 1 ps to assume its new stable orientation, which means that the driven rotation can be considered quite close to simultaneous.

The total energy curve of the rotations along the PF$_3$ chain (shown in Fig. 2) reveals similar minima at 30˚ and 90˚ rotation, as expected from symmetry. This has been confirmed by fully unconstrained optimizations of the PF$_3$ chain with initial driver rotations of 30˚ vs. 90˚ yielding almost equal total energies (the two-step approach slightly breaks the 3-fold symmetry, so that the results at e.g. 30˚ and 90˚ are slightly different). As shown in Fig. 2, configurations around 30˚ and 90˚ are most stable, which is understood by the dominance of PF$_3$ – PF$_3$ repulsion. Between two minima the system exhibits a potential barrier of nearly 2 kcal/mol (0.1eV). According to the total energy curve, the gears in Fig. 1 for 0˚ can spontaneously turn to the 30˚ geometry, but kT at room temperature (about 0.59 kcal/mol or 0.026 eV) is insufficient to climb the barrier from 30˚ to 60˚ without external forces (and any rotational energy picked up from 0˚ to 30˚ would likely be mostly dissipated in the system).

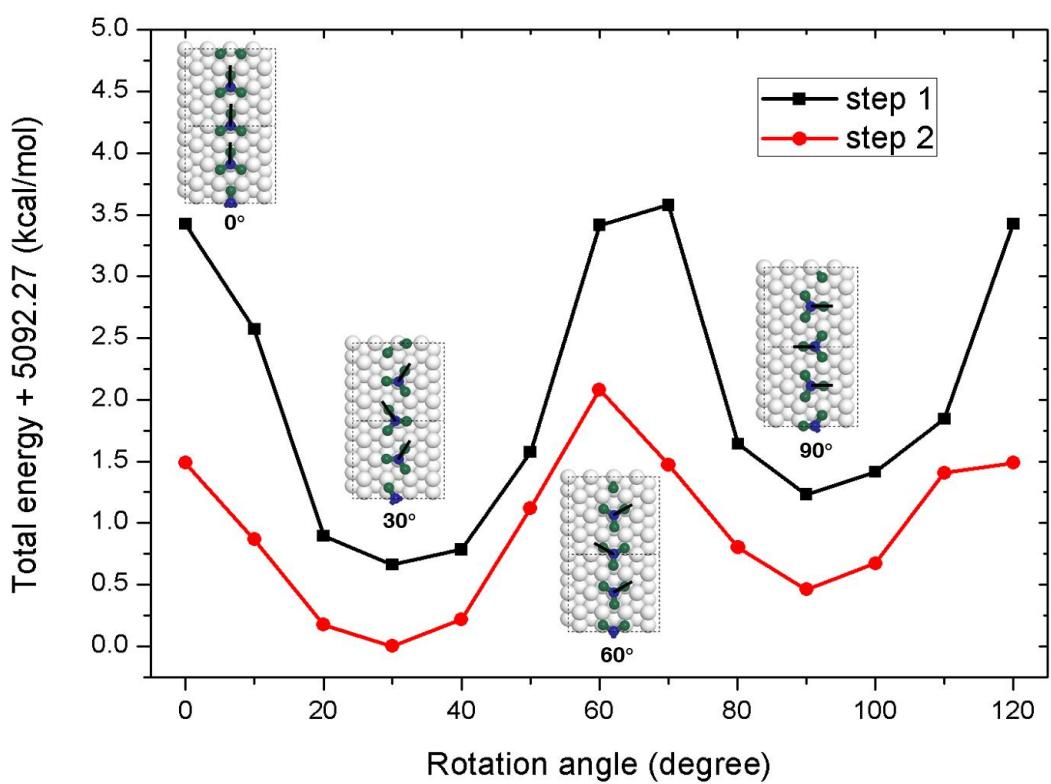

**Fig. 2** Total energy of the chain of PF$_3$ gears of Fig. 1. The rotation angles correspond to those indicated in Fig. 1. As detailed in the text, the simulation at each rotation angle contains two steps: (1) rotate the driver molecules clockwise by 10˚ while keeping the slave molecules fixed (yielding the black curve "step 1"); (2) re-optimize the slave molecules while fixing the drivers at their 10˚ rotation geometry (yielding the red curve "step 2").

For comparison, the NH$_3$ molecule, substituted for PF$_3$ in the chain of molecular gears, has a similar geometry but a weaker polarization than PF$_3$, which weakens the interaction between neighboring NH$_3$ molecules (see also Section II of the Supplementary Information). As we will analyze further below, cf. Table 1, the interaction strength for NH$_3$ is only 20% of that between PF$_3$ molecules at the same intermolecular distance. Thus, NH$_3$ molecules would function less well as meshing gears at the intermolecular distance $d_c = \sqrt{3}d$. As shown in Fig. S1, the relatively short gear arms easily slip past each other; the only smaller distance available on the Cu(111) surface is $d_c = d$, which is however too small for adjacent NH₃ molecules. As shown in Table 1 of the next section, both intermolecular van der Waals attraction and electrostatic repulsion are much weaker for the NH₃ chain than for the PF₃ chain. The weaker H-H (vs. F-F) repulsion is the main reason, while the relatively larger intermolecular distance contributes to a lesser extent.

**Strength of the intermolecular interaction.** As mentioned, we consider only intermolecular interactions resulting from slow quasi-static rotation of the driver molecules while ignoring dynamic effects. The strength of the driver-slave coupling and, thus, the capability to rotate also depend on the intermolecular distance along the PF₃ (and NH₃) chain. Keeping the on-top site adsorption of the molecules at the Cu(111) surface we can also consider chains with intermolecular distances $d_c = d$ and $d_c = 2d$ compared with the initial distance $d_c = \sqrt{3}d$. In both cases, the PF₃ molecules fail to work as gears. For distances fixed at $d_c = d$ the repulsive interaction between adjacent molecules is too strong, exceeding even the adsorbate-substrate interaction, and the PF₃ molecules will be expelled from the chain. In contrast, for distances fixed at $d_c = 2d$ the intermolecular interaction is too weak, such that the slaves cannot follow the driver rotation and thus slip as NH₃ does in Fig. S1.

To analyze more thoroughly the strength of the interaction between adjacent gears, we adopt the energy decomposition scheme of Morokuma, Ziegler and Rauk,(*27, 31*) which is implemented in the ADF package. Here the total interaction energy $E_{int}$ is decomposed according to

$$E_{int} = E_C + E_{Pauli} + E_{orb} + E_{disp} = E_{steric} + E_{orb} + E_{disp} \qquad (1)$$

Here $E_C$ denotes the classical Coulomb interaction between the unperturbed charge distributions of the two molecules and $E_{Pauli}$ represents the Pauli repulsion caused by non-orthogonality of the occupied component orbitals. The sum of these two terms, $E_{steric}$, is usually called steric repulsion (also frozen orbital interaction(*32*)). The orbital interaction energy $E_{orb}$ is due to charge transfer, polarization, and covalent bond formation, while $E_{disp}$ accounts for the dispersion energy resulting from intermolecular van der Waals coupling.

|  | Orientation | $E_{steric}$ | $E_{orb}$ | $E_{disp}$ | $E_{int}$(kcal/mol) |
|---|---|---|---|---|---|
| **PF₃ - PF₃** | 0° | 2.12 | -0.45 | -0.94 | 0.73 |
| **PF₃ - PF₃** | 30° | 0.73 | -0.20 | -0.89 | -0.36 |
| **PF₃ - PF₃** | 60° | 2.27 | -0.43 | -0.94 | 0.90 |
| **PF₃ - PF₃** | 90° | 0.90 | -0.20 | -0.90 | -0.20 |
| **NH₃ - NH₃** | 0° | 0.40 | -0.05 | -0.21 | 0.14 |

**Table 1** Energy decomposition for the interaction energy between two PF₃ or NH₃ molecules without substrate, see Eq. 1. The steric interaction includes both the Coulomb and Pauli interactions. The distance between the centers of two adjacent molecules (viz. phosphorus or nitrogen) was set at $d_c = \sqrt{3}d$, corresponding to the molecular geometry at the Cu(111) surface shown in Fig. 1.

As Table 1 shows, the intermolecular interaction between adjacent PF₃ gears at a separation of $d_c = \sqrt{3}d$ is much stronger than that between NH₃ gears at the same intermolecular distance. The main driving force comes from steric repulsion and van der Waals attraction, the proportion of which depends strongly on the shape – and therefore also orientation – of the molecule. The dispersion interaction turns out to depend only weakly on the gear orientation, whereas the steric repulsion changes markedly with the rotation of the molecules: for the orientations of 0° and 60° (in Fig. 1) it is much stronger than for 30° and 90°. As a result, the total interactions for orientations 30° and 90° are much smaller than those for 0° and 60°, which also reflects the difficulty in climbing the potential barrier from 30° to 60° and the ease of descending along the total energy curve from 60° to 90° shown in Fig. 2.

The molecular motion of the PF$_3$ gears can be efficiently analyzed in more detail by examining the rotational torques experienced by the individual atoms and the total torque on each slave gear. The application to molecules of the classical-mechanics concept of torque is analogous to the common use of quantum-based forces and Newtonian mechanics in ab initio geometric optimization and in molecular dynamics(*33*). The effective total torque $\vec{T}^S_{eff}$, which measures the rotational force applied to any given cluster of atoms (such as gears), is given by(*33*)

$$\vec{T}^S_{eff} = \vec{T}^S_{tot} - \vec{T}^S_{trans}, \tag{2}$$

with

$$\vec{T}^S_{tot} = \sum_i^S (\vec{r}_i - \vec{R}_{ref}) \times \vec{F}_i, \quad \vec{T}^S_{trans} = (\vec{R}_{CoM} - \vec{R}_{ref}) \times \vec{F}^S_{tot}, \tag{2a}$$

$$\vec{F}^S_{tot} = \sum_i^S \vec{F}_i. \tag{2b}$$

Here $\vec{T}^S_{tot}$ is the total torque acting on the cluster (with respect to a reference center $R_{ref}$) given by the sum over all its atoms at positions $\vec{r}_i$ experiencing forces $\vec{F}_i$. Further, $\vec{T}^S_{trans}$ is the torque contribution due to the translational motion of the center of mass (CoM) of the gear with respect to $\vec{R}_{ref}$, where $\vec{R}_{CoM}$ denotes the CoM position and $\vec{F}^S_{tot}$ is the total force acting on the gear. This correction term enters eq. (2) in order to isolate the purely rotational motion to be represented by $\vec{T}^S_{eff}$ from the component of the translational motion of the CoM included in $\vec{T}^S_{tot}$. All forces $\vec{F}_i$ in (2a) and (2b) are obtained as Pulay forces(*34*) for the individual optimized cluster structures by separate single point DFT calculations. For the PF$_3$ gear, the reference center $\vec{R}_{ref}$ for the gear rotation may be assumed to coincide with the phosphorus nucleus. However, the effective total torque $\vec{T}^S_{eff}$ defined by (2), (2a) and (2b) does not depend on $\vec{R}_{ref}$ but only on the position of the gear CoM since simple calculus shows that

$$\vec{T}^S_{eff} = \sum_i^S (\vec{r}_i - \vec{R}_{CoM}) \times \vec{F}_i, \tag{2c}$$

which describes the total torque acting on the cluster with respect to the CoM.

The effective total torque on the PF$_3$ slave at the Cu(111) surface is calculated for rotation angles of the driver between 0° and 120°, using forces obtained from the step 1 calculations (rotated driver for fixed slave), as described above. The numerical analysis shows that for all angles the direction of the slave torque is tilted only slightly with respect to the Cu(111) surface normal, with minor lateral shifts of the slave P atom parallel to the surface (showing also that lateral diffusion of the molecules along the surface is energetically very unfavorable). Therefore, it is sufficient to consider the torque component perpendicular to the surface. As a result, Fig. 3 shows the projected effective total torque on the $PF_3$ slave together with the individual torque components on its fluorine atoms (labeled A, B, C in Figs. 1 and 3).

The total effective torque is found to be always positive, indicating that the rotational force on the slave is unidirectional, with the slave rotating anti-clockwise against the driver $PF_3$ (which itself rotates clockwise). Unidirectionality of rotation is particularly important in any kind of useful machinery. The slave torque increases steeply between 0° and 10° and then decreases until 60°. This initial increase can be understood by the gear geometry near 0°, see Fig. 1, as fluorine atom A of the slave gets close to the adjacent driver, causing largest gear-gear coupling. This coupling is weakened for larger angles when the F arms of the adjacent gear move away from each other, which explains the torque decrease until 60°. At 60° fluorine atom B of the slave gets close to the adjacent driver, analogous to atom A for 0°, thereby starting a repeat of the rotation from 0° to 60° (since the 60° geometry is the mirrored equivalent of the 0° geometry, except for the minor effect of the asymmetry of deeper Cu layers in the fcc stacking sequence), with the atom permutation A→B→C→A. After driver orientation 120°, the whole sequence is repeated identically, because of the three-fold symmetry of both the molecules and the substrate, and this of course repeats again after 240° and 360°.

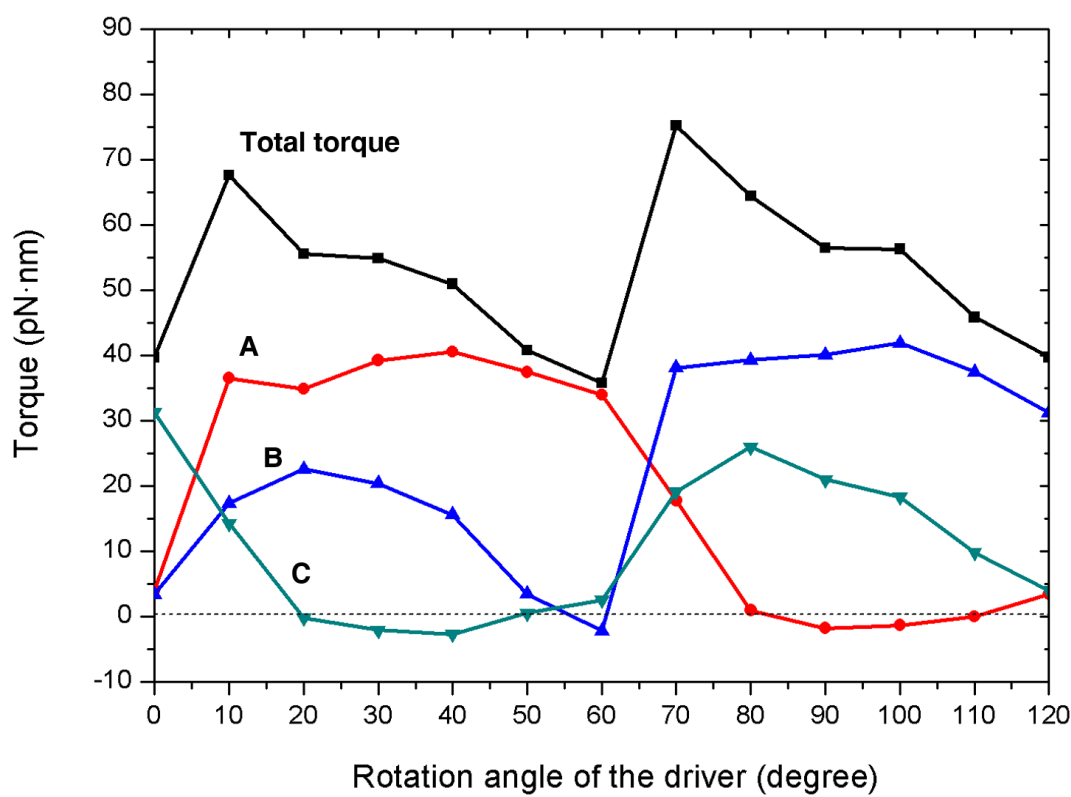

**Fig. 3** Projected effective total torque on the $PF_3$ slave together with its fluorine atom components where the atoms are labeled A, B, C according to Fig. 1, see text. Positive (negative) torque values correspond to counterclockwise (clockwise) rotation along the surface normal.

The individual atom torques show also an interesting qualitative behavior reflecting the gear symmetry. Between 0° and 60° fluorine atoms A and B experience a similar positive torque while the torque on atom C becomes quite small. This can be understood by the geometric arrangement of the fluorine atoms, see Fig. 1: with increasing angle atom C moves away from its nearest driver $PF_3$, which causes a possible torque, while atoms A and B remain between the fluorine arms of the two adjacent drivers. Further, the variation of the torque on fluorine atom A between 0° and 60° resembles that for atom B between 60° and 120°. Likewise, the torque variation of atoms B and C between 0° and 60° is similar to that of atom C and A, respectively, between 60° and 120°. This cyclic behavior of the different atom torques results from the same A→B→C→A permutation described above, due to the three-fold symmetry inside the gear.

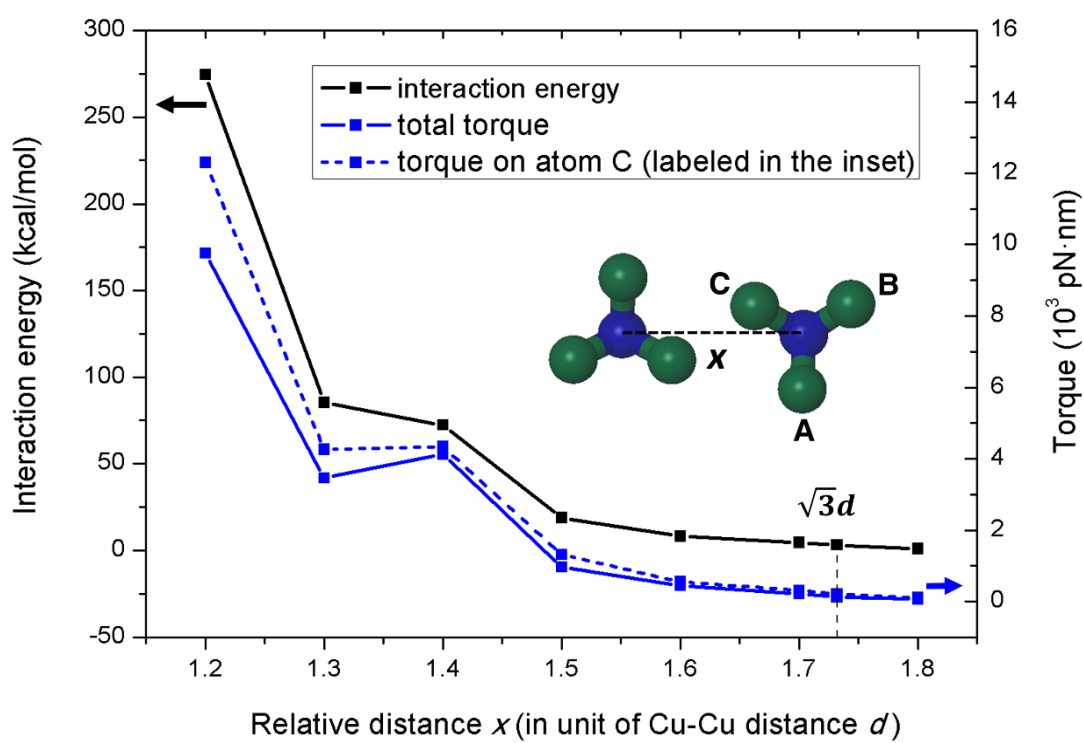

**Fig. 4** Interaction energy (black curve) between two free-standing $PF_3$ molecules (driver and slave gears) with fixed parallel orientation at a variable distance $xd$. The distance is given in multiples $x$ of the nearest neighbor distance $d$ at the Cu(111) surface ($d \approx 2.556$Å) The figure includes the total torque (full blue curve) on the slave gear and the individual torque (dashed blue curve) on its fluorine atom C, see inset.

The intermolecular distance is an important factor for the interaction strength, although we cannot adjust it to continuous values given the discrete adsorption sites available on the substrate (some spontaneous lateral shifting of molecules is possible, but such shifting is energetically very unfavorable compared with the rotation barriers shown in Fig. 2 and can only produce molecular tilting). Choosing different high-symmetry adsorption sites or a metal with a different lattice constant leads to different structures and the meshing between the gears may become more (or less) efficient. This information should be useful for designing molecular gears. To explore this aspect, we plot the interaction energy and the total effective torque for two free-standing parallel PF$_3$ molecules at different intermolecular distances in Fig. 4. Clearly, as the distance between the two gears becomes smaller, the driving force (reflected by the magnitude of the torque) increases simultaneously with the interaction energy. Further, the effective total torque on the PF$_3$ slave molecule is dominated by the torque acting on its fluorine atom C which is closest to the driver molecule, as expected.

In exploring molecular gears, we wish to learn how large the driving force can be, and one definite rule is: the largest distributed driving force that a molecular gear can bear (or transfer) should be no larger than the bond strength between the atoms it consists of; otherwise the gear will be severely distorted or even broken. Also, the force must not be so large that it pushes the molecule off its adsorption site or tilts it out of the way of the driver molecule. That is why the PF$_3$ molecules cannot adsorb on the copper surface with an intermolecular distance of $d_c = d$ (the copper-copper atom distance): optimization shows that the PF$_3$ molecules on every other site will be expelled. For real applications it is thus necessary to adjust the distance between the two gears by choosing a suitable metal with a proper adsorption site allowing appropriate intermolecular distances.

**Multi-arm molecular gears based on carbon rings.** The largest shortcoming of PF$_3$ and its analogues in acting as molecular gears is that they only have three cogs/arms, which renders the interlocking between gears too loose: unlike the case of hard metallic teeth in macroscopic gears, slippage is favored by the wide 120° gaps between their three arms, by the inherent softness of the arms, as well as by the thermal vibrations. Most real applications will need more tightly coupled gears, especially if a string or array of many slave gears need to be driven (we address these cases in a forthcoming publication(*21*)). This requires: (1) a strong bond to the substrate to withstand strong intermolecular forces and (2) gears with many arms/cogs for tight coupling to other gears.

We can generate gears with more arms by using 5-membered carbon rings extended with appropriate chemical groups (such as cyano and ethynyl groups), but such gears are not likely to bind strongly and turn freely on a metal substrate. A better choice is to use a graphene surface with an intermediate metal atom as a supporter and pivot, since metal atoms can be easily adsorbed on graphene.(*35*) It has been reported that a large 5-arm molecular rotor on a metal pivot has the potential to act as a gear.(*36*) In the present work, by choosing a suitable metal atom, we show that it is possible to fix carbon rings on a graphene sheet (or a graphite surface), forming a sandwich structure with the interstitial metal atom.

Sandwich compounds have been studied in organometallic chemistry for several decades.(*37-42*) In the present work, we highlight that manganese and its congeners in the same column of the Periodic Table appear to be suitable intermediates for a 5-membered carbon ring to stand firmly but rotate freely (the rotational barrier is less than 0.005 eV) on a graphene surface with significant barriers (see below for the binding energy) to lateral diffusion along the surface. The bonding mechanism is similar to that of ferrocene, a very stable chemical compound. In a ferrocene molecule, the iron atom in the sandwich structure can donate 1 electron to the two adjacent 5-membered rings, thus making them aromatic. The 12 electrons (6 from each ring) are then shared with the iron atom via covalent bonding so that its outer shell electrons form an 18-electron configuration that is chemically stable. This mechanism holds for similar sandwich structures made of 5-membered rings and 6-membered rings with the prerequisite of having an appropriate metal atom in between, as illustrated in Fig. 5. Such sandwich structures have been synthesized, e.g.

benzene(cyclopentadienyl)manganese(I)(*43*) ($C_6H_6MnC_5H_5$) and ($\eta^6$-graphene)$_2$Cr(*44-46*).

Binding energies of all sandwich structures designed on graphene sheets were calculated using VASP and shown to be rather robust and comparable in structure with that of ferrocene, as listed in Table 2. Here the binding energy is defined as

$$\Delta E = E_{total} - E_{pivot} - E_{rotor} - E_{sub}, \tag{3}$$

where $E_{total}$ is the total energy of the supercell containing one layer of graphene and one rotor molecule consisting of the carbon ring and the intermediate metal pivot atom; $E_{pivot}$ and $E_{rotor}$ denote the total energies of an isolated metal atom and the isolated rotor molecule, respectively; $E_{sub}$ is the total energy of the isolated substrate, viz. the $8 \times 8$ graphene supercell (containing 128 carbon atoms). All energies were obtained from relaxed geometries.

| System | $d_1$(Å) | $d_2$(Å) | $\Delta E$ (kcal/mol) |
|---|---|---|---|
| $C_5H_5 \cdot Fe \cdot C_5H_5$ (Ferrocene) | 1.63 | 1.63 | -289.64 |
| $C_5(CN)_5 \cdot Mn \cdot$ Graphene | 1.63 | 1.62 | -250.44 |
| $C_5(CN)_5 \cdot Tc \cdot$ Graphene | 1.78 | 1.78 | -265.43 |
| $C_5(CN)_5 \cdot Re \cdot$ Graphene | 1.78 | 1.79 | -274.42 |

**Table 2** Interplanar distances and binding energies of several sandwich structures. $d_1$ denotes the perpendicular distance between the carbon ring plane and the metal atom, while for structures containing graphene, $d_2$ denotes that between the metal atom and the graphene plane.

Considering the 5-membered carbon rings with 5 cyano groups above manganese atoms adsorbed on graphene, we explore linear chains with alternating driver and slave molecules analogous to the PF$_3$ gears: in the simulations shown in Fig. 5, we rotate the drivers successively by 10˚ clockwise until 70˚; after 72˚ (1/5 of a full turn) the cycle repeats. Results for the orientations 0˚, 30˚ and 60˚ are depicted in Fig. 5. Clearly, the gears, which are separated by 4 graphene lattice constants, rotate together well: the slave gears can closely follow the drivers.

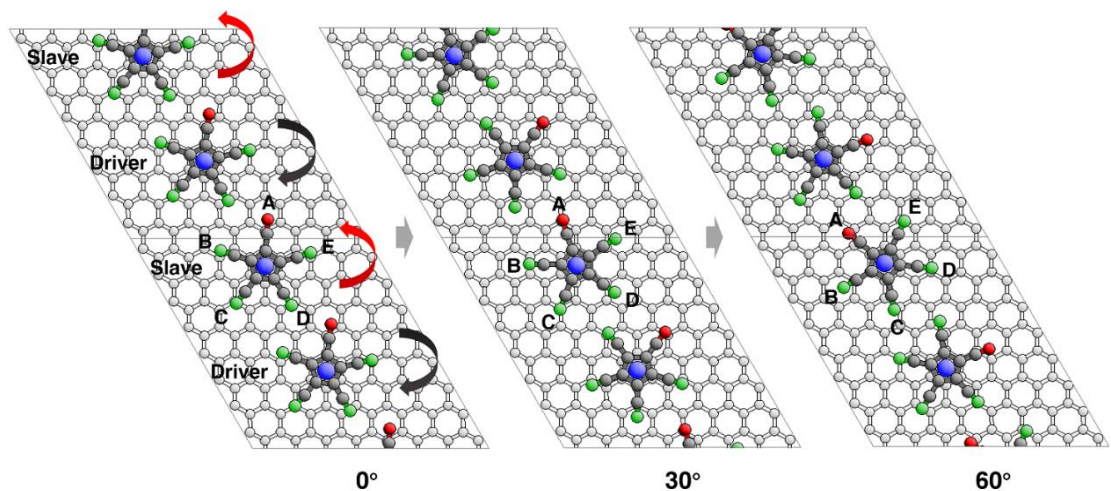

**Fig. 5** Periodic linear chain of molecular gears consisting of 5-membered carbon rings with cyano (-CN) groups (one driver and one slave per unit cell), analogous to Fig. 1. Each gear is attached to graphene with a central manganese atom (blue) acting as the pivot. The intermolecular distance is equal to 4 graphene lattice constants. The CN arms of a slave rotor molecule are labelled A-E. The curved red arrows illustrate their (counterclockwise) rotation while black arrows indicate the (clockwise) rotation of the driver molecules. All molecules have one of their nitrogen atoms highlighted in red (instead of green otherwise) to illustrate the rotations. The three panels correspond to three orientations of the driver molecules: 0°, 30° and 60°, respectively.

The assumption of a linear chain of gears, as used in all examples above, can be restrictive. An alternative option, a zigzag alignment as illustrated in Fig. 6, provides more flexibility since the lateral off-set from the linear alignment is adjustable. As a result, each gear is now relatively free to adjust its distance to the neighboring gears, by tilting and/or sideways sliding, whether inward (toward a more linear alignment) or outward (toward a less linear alignment). However, sideways sliding may reduce the bond strength to the support and may favor tilting of the rotation axis, both of which increase the risk of slipping and skipping of gear arms. In the example of Fig. 6, it can be seen that the slave $C_5$ rings indeed tilt and slide somewhat off-site due to their mutual repulsion.

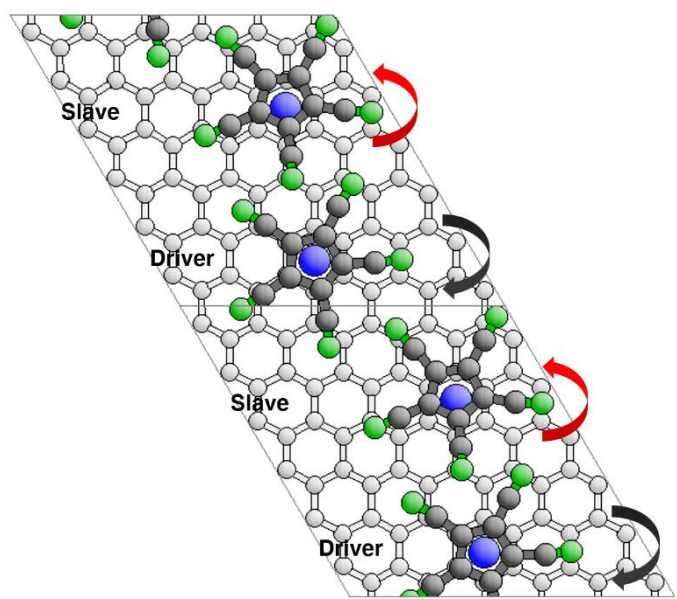

Fig. 6 Zigzag alignment of molecular gears of 5-membered carbon rings with cyano (-CN) groups and manganese pivots.

**Criteria for designing interlocking molecular gears**

Based on our results for a series of different gear/support systems, we can identify several important general criteria for designing a successful interlocking of molecular gears, i.e. for ensuring that the rotation of one gear molecule is transmitted effectively to a neighboring gear molecule. It is interesting to note that a quasi-classical picture of these rotations is realistic at least at a qualitative level, while quantum effects clearly underlie these motions at the quantitative level; also, special quantum effects such as tunneling are not considered here; furthermore, adding thermal effects and energy dissipation will also have quantitative consequences. While our driver-slave model systems are by no means exhaustive of all possible molecular gear systems, we believe that the following individual criteria will apply to many instances of "mechanical" (i.e. non-bonding) contacts between surface-supported molecules:

*1) Stiffness of support*: The support (e.g. a surface) must hold the gears at optimal gear-gear distances and must keep the gear rotation axes well oriented (e.g. parallel to each other). The support should be stiff enough to bear the strain of the structural distortion, usually caused by the interaction between the gears it is supporting. For the cases shown in Fig. 5, a closer distance between two gears (such as two graphene lattice constants apart instead of four) can cause the graphene surface to be torn up, which means that the strain exceeds the threshold that the C-C bonds can bear. Our theoretical optimization of hexacyanobenzene (benzene substituted with 6 cyano groups) also shows that when this kind of gears is aligned with a very small separation, the graphene surface will be broken up. However, we suggest that this molecule may exhibit good performance on a more robust surface such as some metal surfaces or a graphene monolayer adsorbed on a metal surface(*47*).

*2) Strength of attachment of gears to support*: Strong attachment to the support is desired, but must not be so strong as to break up the support (or the gear itself), as mentioned in criterion 1). Also, gears should not slide away along the surface under the applied rotational forces, i.e. there should be sufficiently large barriers to lateral diffusion.

*3) Easy rotation of gears*: The attachment of a gear to the support should present the smallest possible rotational energy barriers, to allow free rotation, despite the requirement of strong gear-support attachment mentioned in criterion 2). This is why we chose a metal atom as the support on a graphene sheet: the existence of a suitable metal atom (the pivot) greatly reduces the rotational energy barrier.

*4) Stiff gears*: Gears must be stiff enough to prevent bending that would favor slippage and skipping past the arms of meshed gears. A counter-example is the $C_6(C_2H)_6$ gear which we have explored in parallel work on gear chains that consist of one driver and several slaves(*21*): when we add more slave gears (e.g. 3 slave gears with one driver gear), we find that the long arms can slip and skip at some rotation angles.

*5) Multi-arm gears*: The opening angle between successive arms (measured as 360° divided by the number of arms) should be minimized so that arms of meshed gears can grip each other well. This favors multiple arms, which in turn implies a large gear diameter, but it may be difficult to design a gear of large diameter that is sufficiently circular and angularly periodic to match well a meshed gear.

*6) Suitable gear-gear interactions*: We can choose among different intermolecular forces (Coulomb, van der Waals, etc.) to optimize the interaction between gears, by selecting the chemical composition and structure of the gears. Attractive van der Waals forces are strongly distance-dependent and thus limit the range of acceptable gear-gear distances, while Coulomb forces are stronger and have a longer range and thus allow a wider spread of gear-gear distances. On the other hand, Coulomb forces are also associated with more reactive chemistry. Forces due to steric repulsion will dominate in gears with closed shell structure, as in the examples discussed in this work: these are indeed generally dominant in the "mechanical bond" that is fundamental to molecular machines.(*48*) Typically, the gear-gear interaction can be further controlled by adjusting factors such as gear-gear distance, gear shape, and gear alignment, which are discussed next.

    *a) Gear-gear distances*: The repulsive gear-gear interactions decay fast with increasing separation between gears, while close separations increase the risk of gear-gear bond formation that prevents rotation. Also, too small gear-gear separations can make gears tilt out of optimum contact, or expel or destroy gears and/or the support. Finding appropriate gear-gear distances is both crucial and delicate. Thus, hexafluorobenzene (benzene substituted with 6 fluorine atoms) failed to act

as molecular gears in some of our simulations on graphene, because its arms were too short for the intermolecular distance used; however, a closer distance (two graphene lattice constants apart) leads to the aforementioned gear-gear bond formation. Choosing a different supporting surface (e.g. a metal surface) with another lattice constant provides further opportunities to adjust the gear-gear separation.

*b) Gear alignment*: We have mainly considered linear chains of gears in this study, while other arrangements offer other opportunities for adjusting gear-gear interactions. The zigzag alignment described above (cf. Fig. 6) adds at least two degrees of freedom of motion, namely out-of-line shifting and tilting (which in addition may also vary during the rotation itself). Such additional degrees of freedom of motion may not always be beneficial, as they could in some cases allow the gear to escape, especially when strong repulsive forces between gears exist. The "boxing in" provided by a linear alignment of adsorption sites can thus help keep the chain of gears under control. These considerations will also become important with more complex two-dimensional arrays of linked gears, where the geometrical relationship between multiple neighboring gears needs to be optimized. The exact details will certainly depend on the specific molecules and substrates used, so that the ability to optimize the choice of materials becomes very beneficial to designing the best possible gear system.

*c) Matching gear arm shapes*: Arms in meshing gears should match well at all rotation angles (gear arms should fit well inside the grooves between opposing arms). With identical gears, the arm shape must therefore be "self-complementary", as any intentional arm shape change on one gear will also appear on the arms of the other gear in a complementary fashion. In particular, the wiggle room between arms on meshing gears must be minimized to allow effective transmission of rotation with minimal angular delay (this also depends on the gear-gear distance and the gear-gear interactions). One option to achieve better gear-gear matching would be to use alternating different gears, i.e. two different kinds of molecules: such mixing opens up a wide range of interesting opportunities, including the use of molecules with alternating electric charges that favor self-assembly in alternating one- or two-dimensional arrays.

## Conclusions

The present work investigates several prototypes of molecular gears, from triple-arm gears ($PF_3$ or $NH_3$ molecules attached to copper surfaces) to multi-arm gears (a 5-membered carbon ring substituted with 5 -CN groups attached to graphene sheets).

By exploring the "soft" interlocking mechanism and corresponding intermolecular interactions, we reveal the advantages and disadvantages of different choices of gears and substrate, and can propose a series of general criteria for designing molecular gears. These criteria are not always mutually compatible, so compromises are usually required. Our results presented here give initial indications of potentially successful gear compositions and architectures. We find that multi-arm Coulomb-dominated gears operate well on a metal surface, while multi-arm C-ring-derived gears can be successful on a graphene sheet. In parallel work, we focus on more distant transmission and explore longer finite gear chains containing one driver and several slaves: we find that transmission of rotation beyond a few slave gears is more sensitive to these effects.(*21*) Similar structures on other suitable substrates are expected to also show such gear effects and these molecular gears are expected to be helpful for designing larger cooperative molecular machines and networks.

**Acknowledgements**

This work was supported by the Collaborative Research Fund of Hong Kong Research Grants Council (Project No. C2014-15G). We also acknowledge the computing resources of the Tianhe2-JK cluster at the Beijing Computational Science Research Center and of the Tianhe2 cluster at the National Supercomputer Center in Guangzhou, China. K. Hermann is grateful for support by the Institute of Computational and Theoretical Studies (ICTS) at HKBU; ICTS is supported by the Institute of Creativity, which is sponsored by Hung Hin Shiu Charitable Foundation (孔憲紹慈善基金贊助).

Supplementary Materials

**Interlocking mechanism between molecular gears attached to surfaces**

Rundong Zhao[1], Yan-Ling Zhao[2], Fei Qi[1], Klaus Hermann[3], Rui-Qin Zhang[2,4] and Michel A. Van Hove[1*]

[1]Institute of Computational and Theoretical Studies & Department of Physics, Hong Kong Baptist University, Hong Kong SAR, China
[2]Department of Physics and Materials Science, City University of Hong Kong, Hong Kong SAR, China.
[3]Inorganic Chemistry Department, Fritz-Haber-Institute der Max-Planck-Gesellschaft, Faradayweg 4-6, 14195 Berlin, Germany
[4]Beijing Computational Science Research Center, Beijing, China

* Corresponding author; E-mail: vanhove@hkbu.edu.hk

**I. Simulation of the NH$_3$ gears analogous to the PF$_3$ gears.**

Calculations analogous to the case of PF$_3$ on Cu(111) are performed for a linear chain of NH$_3$ molecules on the same Cu(111) surface with alternating drivers and slaves, positioned with the same geometry as discussed for the PF$_3$ chain. They show a rotational behavior along the chain which is somewhat different from that of PF$_3$. For (clockwise) driver rotations up to 60° the slaves follow the rotation (anti-clockwise) with some delay, whereas beyond 60° until 90° the slaves move by very little: they slip. For 90°-120°, they barely turn any more, rotating less than 5°, and do not return to their initial orientation. This can be understood by the smaller arms/cogs of NH$_3$ (N-H arms) as compared with those of PF$_3$ (P-F arms) which weaken rotational coupling and facilitate slippage, especially at this relatively large intermolecular distance compared to the PF$_3$ case. It is also supported by the total energy dependence on rotations along the NH$_3$ chain (not shown): it is analogous to that of Fig. 2 which exhibits a similar overall shape, however, with a rather small potential barrier near 60° (less than 0.01eV, i.e. ~0.2 kcal/mol); in fact, this barrier height evaluated within the DFT framework is too small to be numerically reliable. Still, the shape of the total energy curve for NH$_3$ is very similar to that of Fig. 2 and can explain the failure of NH$_3$ to co-rotate fully: there is a potential barrier from 30° to 60° but the interaction force is too weak to push the system over this barrier. At the next closer intermolecular separation, the Cu-Cu distance *d*, the NH$_3$ molecules are too tightly packed and are expelled from the gear chain.

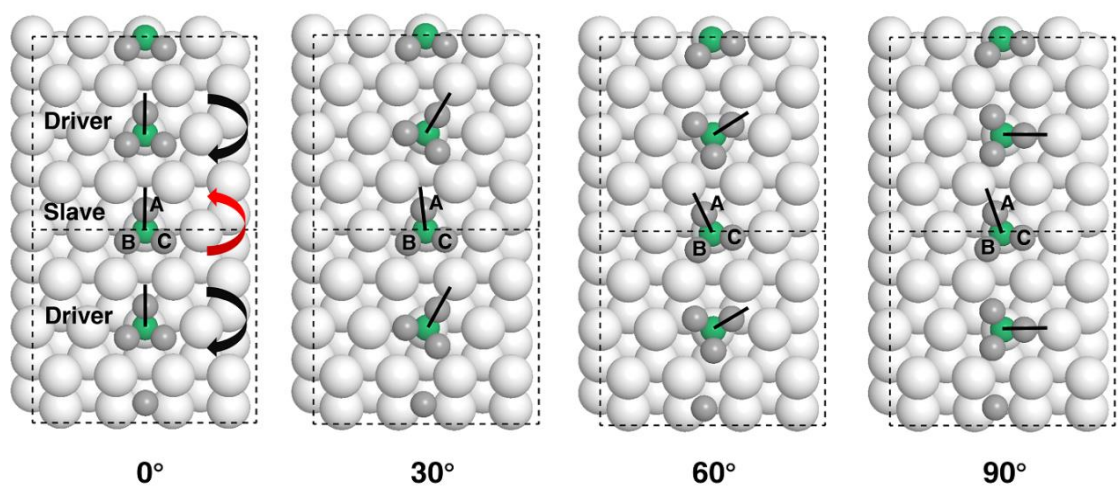

**Fig. S1** Simulation of the rotation in a linear chain of NH$_3$ gears on Cu(111), analogous to the case of PF$_3$ (cf. Fig. 1 in the main text for details of the AB$_3$ type gears).

**II. The "2-step" process.**

The 2-step procedure is illustrated in detail in Fig. S2. We start from the geometry given in Fig. 1 for a driver and slave orientation of 0°. In order to obtain the geometry of a driver orientation of 10° (black arrows in left panel) we fix in **s**tep 1 the entire slave gear at its initial 0° orientation. Then we perform a single point calculation using this structure and obtain the total force on atoms A, B and C, from which we derive the torques acting on each F atom (the force and torque on the axial P atom is negligible). In **s**tep 2, we re-optimize the slave (red arrow in the second panel from left) while keeping the driver fixed at its 10° orientation − the coordinates of all the atoms in the driver are fixed in the simulation. All the forces and torques acting on atoms A, B and C should vanish after the optimization of step 2. Similarly, in the geometry for the 40° orientation of the driver (the third panel from left), the slave gear actually has the geometry optimized for a driver orientation fixed at 30°.

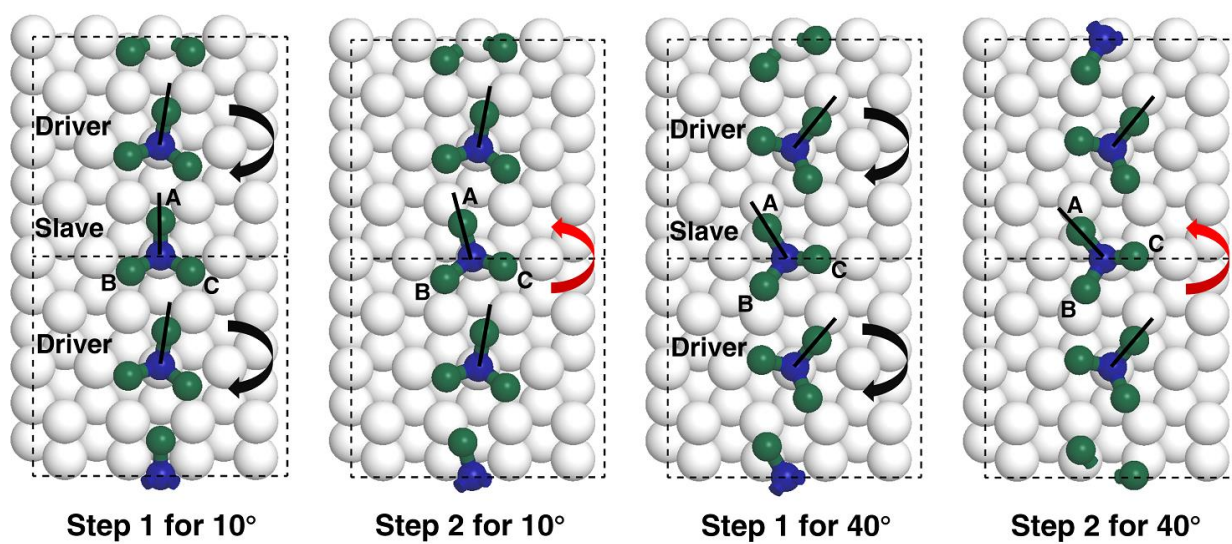

**Fig. S2** Illustration of the 2-step process leading to the results shown in Fig. 2 and 3.